\documentclass[a4paper,10pt,twoside,twocolumn]{article}
\usepackage{amsmath}
\usepackage{graphicx}

\begin{document}

\title{{\vspace{-0.0cm}\sc{Extended mean flow analysis \\
of the circular cylinder flow}}}
\author{Olivier Marquet$$ \& Marco Carini$$\vspace{0.3cm}\\
\small{\it{$$ONERA - Paris Saclay University, Meudon, France}}\\
}
\date{}
\maketitle
\vspace{0.5cm}

\section*{Abstract}
A new eigenvalue analysis is developed and applied to the circular cylinder laminar flow configuration to investigate the various mechanisms at play in the nonlinear saturation of perturbations yielding to limit cycles for supercritical values of the Reynolds number. Unlike the mean-flow analysis, which only accounts for the interaction of the first-harmonic of the time-periodic flow with its mean-flow, the so-called extended mean-flow analysis also accounts for its interaction with the second-harmonic. The results reveal the existence of two eigenvalues both having a growth rate exactly equal to zero. The high-frequency eigenmode gives better estimations of the frequency and spatial structure of the first-harmonic than the marginal modes obtained with the mean-flow analysis, especially when the Reynolds number is increased.

\section{Introduction}\label{sec:intro}
Self-sustained oscillations occur in open flows when an infinitesimal global perturbation gets unstable and then saturates
to a limit-cycle state. Although the linear stability analysis of the underlying steady base flow provides a rigorous mathematical 
description of the initial instability growth mechanisms, its predictive capabilities rapidly reduce as flow oscillations grow 
in amplitude away from the bifurcation. Corresponding leading-order nonlinear effects can be taken into account by introducing a 
weakly nonlinear formalism, which however remains still limited by the perturbative nature of the involved approximation. A classical 
example is represented by the onset of the B\'{e}rnard-von K\'{a}rm\'{a}n vortex street in the wake of a circular 
cylinder\cite{mathis-provansal-boyer-1984,provansal-mathis-boyer-1987,marquet-sipp-jacquin-2008} at Reynolds number $\textit{Re} \approx47$. For this flow, 
the frequency drift from the linear stability estimate\cite{barkley-2006,sipp-lebedev-2007} rapidly increases up to $30\%$ at $\textit{Re}=80$, 
while the error on the oscillation amplitude computed by a weakly nonlinear analysis\cite{mantic_lugo-arratia-gallaire-2014} exceeds $100\%$, with the 
convergence radius of the corresponding normal-form series being vanishingly small\cite{carini-auteri-giannetti-2015}.  

The possibility to better account for the nonlinear effects and then for the saturation mechanism of the flow oscillations, obtaining an 
accurate prediction of their frequency and spatial structure, has moved the attention of the researchers to the stability properties 
of the resulting time-averaged mean flow. Indeed for the laminar cylinder wake\cite{barkley-2006} and other different flow 
configurations, either in the turbulent or laminar regime\cite{crouch-garbaruk-magidov-2007,gudmundsson-colonius-2011,turton-tukerman-berkley-2015}, the 
linear stability analysis heuristically performed on the top of the mean flow yields a mildly unstable eigenvalue at approximately the same frequency 
of the nonlinear unsteady flow, a result which is reminiscent of the seminal works by Malkus\cite{malkus-1956} and Stuart\cite{stuart-1971}. These authors 
first emphasized the role of the mean-flow distortion in the nonlinear saturation of linear perturbations. As disturbances, 
initially fed on the unstable base flow, linearly grow, reaching a finite-size amplitude, the resulting Reynolds stresses induce a 
distortion of the mean flow, which in turn reduces the temporal growth rate of the perturbations until the growth-rate is nearly zero. This conjecture leads 
to invoke the marginal stability property of the mean flow, for which the name `real-zero imaginary frequency' (RZIF) has been introduced 
by Turton, Tuckerman and Barkley\cite{turton-tukerman-berkley-2015}, showing that it exactly holds for pure monochromatic flow oscillations, 
and it is approximately satisfied when higher-order harmonics becomes negligible compared to the leading one, such as in the case of travelling 
waves in thermosolutal convection.    
   
The RZIF property has been exploited by Manti\v{c}-Lugo, Arratia and Gallaire\cite{mantic_lugo-arratia-gallaire-2014} to derive a 
self-consistent model of the saturation dynamics of the cylinder wake, by formalizing the above described phenomenological picture. In their 
model the nonlinear fluctuation is represented by the mean flow leading global mode, with the mean flow being determined as the solution of 
the steady Navier-Stokes equations forced by the Reynolds stresses resulting from the interaction of the leading mode with itself. The arbitrary 
mode amplitude is computed so that the mean flow distortion yields a zero growth-rate, thus assuming the RZIF property to close the model. The 
authors show that their model accurately predicts all the relevant features of the flow obtained from direct numerical simulations (DNS) up to 
$\textit{Re}=110$, including flow oscillation frequency, amplitude and spatial pattern, as well as the mean flow and the Reynolds stresses. Later this model has 
been employed by Meliga, Boujo and Gallaire\cite{meliga-boujo-gallaire-2016} to perform an adjoint sensitivity analysis of the limit cycle frequency and 
amplitude, in order to identify the wavemaker regions of the nonlinear wake oscillations and the most sensitive regions for passive control design. 
However, despite these remarkable results, almost exclusively obtained for the cylinder wake, the RZIF property does not hold for all the 
flows. Well known counterexamples are represented by the oscillations in an open cavity\cite{sipp-lebedev-2007}, in turbulent wakes\cite{meliga-pujals-serre-2012,mettot-sipp-bezard-2014} and 
and by standing waves in thermosolutal convection\cite{turton-tukerman-berkley-2015}. Indeed the feedback between the mean flow distortion and the 
leading mode amplification is not the unique mechanism explaining the nonlinear saturation but another fundamental mechanism exists, which is 
driven by the production of high-order harmonics. For this enriched dynamics, a strong deviation from the marginal stability of the mean flow is 
often observed, and even the frequency prediction is not always accurate. As an example in the case of an open cavity, Sipp \& Lebedev\cite{sipp-lebedev-2007} 
pointed out the role played by the interaction between the leading global mode and its second harmonic, resulting in an unstable mean flow. 
The aim of this paper is to introduce a new analysis that extends the mean-flow analysis as it takes into account both the effects of the mean flow distortion and the nonlinear interaction with the second harmonic.

The paper is organized as follow. The theoretical formulation is introduced 
in \ref{sec:formulation} including the eigenvalue analysis of the base flow, the mean-flow analysis and the extended mean-flow analysis. This new analysis 
is then applied to the circular cylinder flow configuration and results are  presented 
\ref{sec:results}

\section{Theoretical formulation}\label{sec:formulation}
The dynamics of laminar unsteady hydrodynamic flows in a domain $\Omega$ is governed by the following non-dimensional incompressible Navier-Stokes equations
\begin{equation}\label{eq:nlnsc}
 \mathcal{M} \frac{\partial \mathbf{q}}{\partial t} + \mathcal{N}(\mathbf{q}) = \mathbf{0}, 
\end{equation}
where $\mathbf{q}=(\mathbf{u},p)^{T}$ denotes the flow state composed of the flow velocity field $\mathbf{u}(\mathbf{x},t)$ and the pressure field $p(\mathbf{x},t)$ that depend on the space coordinates $\mathbf{x}$ and on the time variable $t$. The linear operator $\mathcal{M}$ is defined by its action on the flow state as  $\mathcal{M}\mathbf{q}=(\mathbf{u},^0)^{T}$ 
and the nonlinear residual is defined as 
\begin{equation}\label{eq:nsopdef}
 \mathcal{N}(\mathbf{q}) = 
 \begin{pmatrix}
   (\mathbf{u} \cdot \mathbf{\nabla})\mathbf{u} + \mathbf{\nabla}{p} - \textit{Re}^{-1} \mathbf{\Delta} \mathbf{u} \\[2mm]
   \mathbf{\nabla} \cdot \mathbf{u}
 \end{pmatrix},
\end{equation}
where the Reynolds number $\textit{Re} = U_\infty D / \nu$ is based on a reference velocity scale $U_\infty$ and a reference length scale $L$ used to make all the flow variables dimensionless. In the cylinder flow configuration investigated in the next section, they will correspond to the uniform inflow velocity and the cylinder diameter $L=D$. \\
\\
Time-independent solutions of \ref{eq:nlnsc}, often called base flows and denoted hereinafter $\mathbf{q}_{b}=(\mathbf{u}_b,p_b)^{T}$, satisfy the steady non-linear equations
\begin{equation}\label{eq:baseflow}
\mathcal{N}(\mathbf{q}_{b}) = \mathbf{0}. 
\end{equation}
Their temporal stability is investigated by superimposing an infinitesimal perturbation as
\begin{equation}\label{eq:F-pert}
  \mathbf{q}(\mathbf{x},t) = \mathbf{q}_b(\mathbf{x}) + \epsilon  \left( \hat{\mathbf{q}}(\mathbf{x})e^{\left( \sigma_{b} + \rm{i} \omega_{b} \right) t}  + \mbox{ c.c.} \right)  + O(\epsilon^2)
\end{equation}
where $\epsilon$ is an infinitesimally small number and $\mbox{c.c.}$ is an abbreviation of complex conjugate. The linear perturbation is decomposed into a complex-valued spatial structure $\hat{\mathbf{q}}$ growing or decaying exponentially in time at the growth rate $\sigma_{b}$ and oscillating at the frequency $\omega_{b}$. Introducing the above decomposition into 
the non-linear \ref{eq:nlnsc}, using \ref{eq:baseflow} of the base-flow and neglecting higher-order terms $O(\epsilon^2)$, yields the following generalized eigenvalue problem   
\begin{equation}\label{eq:F-eigbaseflow}
\left[ \left(\sigma_{b} + \rm{i} \omega_{b} \right) \mathcal{M} + \mathcal{L}(\mathbf{q}_b) \right] \hat{\mathbf{q}} = \mathbf{0}
\end{equation}
where $\mathcal{L}(\mathbf{q}_b)$ is obtained by linearizing the non-linear residual $\mathcal{N}(\mathbf{q})$ around the base-flow $\mathbf{q}_b$. For the incompressible Navier-Stokes equations, the so-called Jacobian operator is defined as     
\begin{equation}\label{eq:Lns-def}
 \mathcal{L}(\mathbf{q}_b) = 
\left( 
\begin{tabular}{lr}
$(\mathbf{u}_b \cdot \mathbf{\nabla} )+  (\mathbf{\nabla} \mathbf{u}_b) - \textit{Re}^{-1} \mathbf{\Delta} $ &  $\nabla$ \\ [2mm]
 $\mathbf{\nabla} \cdot $ & $0$ 
 \end{tabular} \right)
\end{equation}
The stability of the base-flow is determined by computing the leading eigenvalue, i.e. the eigenvalue having the largest growth rate. When its growth rate is positive ($\sigma_b >0$), the base-flow is unstable. If the frequency of the unstable eigenvalue is different from zero ($\omega_{b} \neq 0$),  the base-flow evolves towards a limit cycle which is an unsteady solution of  \ref{eq:nlnsc} satisfying the periodicity condition
\begin{equation}\label{eqn:periodicity}
\mathbf{q}(\mathbf{x},t+T) = \mathbf{q}(\mathbf{x},t)  
\end{equation}
where $T$ is the period. The frequency of the limit cycle $\omega = 2 \pi / T$ is \textit{a priori} different from the frequency of the unstable mode ($\omega \neq \omega_{b}$), that will be called theireafter the \textit{base-flow frequency}.  The limit-cycle solution is decomposed as 
\begin{eqnarray}\label{eq:F-meanfluct}
  \mathbf{q}(\mathbf{x},t) = \mathbf{q}_0(\mathbf{x}) +  \mathbf{q}'(\mathbf{x},t) 
\end{eqnarray}
the sum of a time-independent component $\mathbf{q}_0(\mathbf{x})$, called the mean-flow, 
and a time-dependent component $\mathbf{q}'(\mathbf{x},t)$ called the fluctuation. This fluctuation is time-periodic and can thus be expanded as a series of harmonics as 
\begin{eqnarray}\label{eq:F-series}
 \mathbf{q}'(\mathbf{x},t) = \left( A\mathbf{q}_1(\mathbf{x})e^{\rm{i} \omega t} + A^2 \mathbf{q}_2(\mathbf{x})e^{2 \rm{i} \omega t} + \cdots \right) + \mbox{ c.c.}  ,\nonumber 
\end{eqnarray}
where $\mathbf{q}_1(\mathbf{x})$ is the (complex-valued) first harmonic oscillating at the fundamental frequency $\omega$ and $\mathbf{q}_2(\mathbf{x})$ is the second harmonic oscillating at twice the fundamental frequency. In the following, higher-harmonics are omitted unless stated otherwise. The first and second harmonics are scaled by the amplitude  $A$ which is a complex number, implictely defined by the following normalization condition 
\begin{equation}\label{eq:normalization}
\int_{\Omega_{s}} \mathbf{q}_1^* (\mathbf{x}) \mathbf{q}_1 (\mathbf{x}) \, d\mathbf{x}  = 1 \, .
\end{equation}
The spatial domain $\Omega_{s}$ is enclosed in the computational domain $\Omega$ and allows a definition of the amplitude $A$ independent of the computational domain. The mean flow, first and second harmonics can be explicitely defined from the knowledge of the instantaneous solution $\mathbf{q}(t)$ over a period $[t_0,t_0+T]$ as 
\begin{equation}\label{eq:F-def}
A^{n} \mathbf{q}_{n} = \frac{1}{T} \int_{t_{0}}^{t_{0}+T} \mathbf{q}(t)  e^{\rm{i} n \omega t} dt \;,
\end{equation}
the instantaneous solution being computed by marching in time \ref{eq:nlnsc} and \ref{eq:nsopdef}.\\
\\
To better characterize the different non-linear mechanisms at play in a limit-cycle, it is interesting to formulate the governing non-linear equations in the frequency domain rather than in the time domain. To that aim, the truncated expansion \ref{eq:F-series} is introduced into \ref{eq:nlnsc}. By balancing the time-independent terms and those oscillating at frequencies $\omega$ and $2\omega$, one obtains the so-called harmonic balanced formulation of the governing non-linear equation \ref{eq:nlnsc}, truncated at the second order,  
\begin{eqnarray}\label{eq:F-eq}
  \mathcal{N}(\mathbf{q}_0)   =  |A|^2 \, \mathcal{C}(\mathbf{q}_1^*)\mathbf{q}_1 &+& |A|^4 \,  \mathcal{C}(\mathbf{q}_2^*)\mathbf{q}_2 ,\label{eq:F-eq:0}\\[2mm]
  \left[ \rm{i}\omega \mathcal{M}  + \mathcal{L}(\mathbf{q}_0) \right] \mathbf{q}_1 &=& |A|^2 \, \mathcal{C}(\mathbf{q}_{2})\mathbf{q}_1^*,\label{eq:firstharmo}\\[2mm]
 \left[ 2 \rm{i} \omega \mathcal{M} + \mathcal{L}(\mathbf{q}_0) \right] \mathbf{q}_2 &=& \frac{1}{2} \, \mathcal{C}(\mathbf{q}_1)\mathbf{q}_1 ,\label{eq:F-eq:2}
\end{eqnarray}
where $(\cdot)^*$ denotes the conjugation operation. The quadratic interaction between an harmonic $\mathbf{q}_n$ and an $\mathbf{q}_m$ is denoted as   
\begin{equation}
 \mathcal{C}(\mathbf{q}_n)\mathbf{q}_m =
\begin{pmatrix}
  -(\mathbf{u}_n \cdot \mathbf{\nabla})\mathbf{u}_m - (\mathbf{u}_m \cdot \mathbf{\nabla})\mathbf{u}_n \\
  0
 \end{pmatrix}
\end{equation}
Note that the sign minus in the above definition. The mean flow is governed by \ref{eq:F-eq:0} similar to \ref{eq:baseflow} that governs the base flow, except for the right-hand side terms that are the quadratic interactions between the harmonics $\mathbf{q}_n$ and their complex conjugate $\mathbf{q}_n^{*}$, where $n=1,2$ because of the second-order expansion. The sum of all these terms is the divergence of the Reynolds-stress tensor. When the amplitude of the fluctuation is zero, i.e. $|A|=0$, it vanishes and the mean-flow equation is exactly the base-flow equation. \ref{eq:firstharmo} and \ref{eq:F-eq:2} govern the first and second harmonics, respectively. The linear operators on the left-hand sides involve the Jacobian operator \textit{Re}f{eq:Lns-def} defined not with the base-flow $\mathbf{q}_b$ but with the mean-flow $\mathbf{q}_0$.  The right-hand side terms couple the two equations.\\
\\
Let us now focus on \ref{eq:firstharmo} that governs the first harmonic $\mathbf{q}_1$. As explained in \cite{turton-tukerman-berkley-2015}, the right-hand side term $|A|^2 \, \mathcal{C}(\mathbf{q}_{2})\mathbf{q}_1^*$ is negligible if the periodic flow is purely monochromatic ($\mathbf{q}_2=\mathbf{0}$) or if the amplitude of the first-harmonic is small ($|A| \ll 1$). In those two cases, \ref{eq:firstharmo} can be approximated by  
\begin{equation}\label{eq:meanapprox}
\left[ \rm{i}\omega\mathcal{M}  + \mathcal{L}(\mathbf{q}_0) \right] \mathbf{q}_1 \approx \mathbf{0} 
\end{equation}
Therefore, when examining the eigenvalue spectrum of the Jacobian operator around the mean flow $\mathcal{L}(\mathbf{q}_0)$ after solving the generalized eigenvalue problem 
\begin{equation}\label{eq:F-eigmeanflow}
\left[ \left(\sigma_{m} + \rm{i} \omega_{m} \right) \mathcal{M} + \mathcal{L}(\mathbf{q}_0) \right] \hat{\mathbf{q}}_{m} = \mathbf{0}
\end{equation}
one should find an eigenvalue whose growth rate $\sigma_{m}$ is close to zero, frequency $\omega_{m}$ is close to the fundamental frequency $\omega$ of the time-periodic flow and associated eigenmode $\hat{\mathbf{q}}_{m}$ approximates the spatial structure of the first harmonic $\mathbf{q}_{1}$. This property of limit cycles is called the Real Zero Imaginary Frequency (RZIF) in \cite{turton-tukerman-berkley-2015} and can be sum up as
\begin{equation}\label{eq:F-properties}
\sigma_{m} \approx 0 \;\;, \;\; \omega_{m} \approx \omega \;\;,\;\; \hat{\mathbf{q}}_{m} \approx \mathbf{q}_1 \;.
\end{equation}
It was first observed in \cite{barkley-2006} for the wake flow behind a circular cylinder flow in the range $ 50 < \textit{Re} < 100$. However, this property of limit cycles is not necessarily satisfied, as shown in \cite{sipp-lebedev-2007} for an open-cavity flow configuration. In that case, the quadratic interaction term $(\mathcal{C}(\mathbf{q}_{2})\mathbf{q}_1^*)$ in \ref{eq:firstharmo} is not negligible. Using a weakly non-linear expansion of the flow to compute the coefficients of the equation governing the temporal evolution of the amplitude $A$, Sipp \& Lebedev \cite{sipp-lebedev-2007} quantified the role of the mean-flow and second-harmonic in the nonlinear saturation of the amplitude. However, the validity of the weakly non-linear analysis is limited to small amplitudes that are reached close the bifurcation threshold. \\
\\
To circumvent such limitation, we introduce a new eigenvalue analysis that allows to account for both effects, unlike the mean-flow analysis. 
The complex-valued first and second harmonics are first decomposed into their real and imaginary part as, 
$\mathbf{q}_{1} = \mathbf{q}_{1}^{r}+ \rm{i} \mathbf{q}_{1}^{i}$  and $\mathbf{q}_{2} = \mathbf{q}_{2}^{r}+ \rm{i} \mathbf{q}_{2}^{i}$. By introducing these decompositions in \ref{eq:firstharmo} and splitting the imaginary and real contributions, one obtains the coupled system of equations
\begin{eqnarray}\label{eq:firstharmorealimag}
  \omega \mathcal{M} \mathbf{q}_1^{r}  + \mathcal{L}\left[\mathbf{q}_0 + |A|^2 \, \mathbf{q}_{2}^{r} \right]  \mathbf{q}_1^{i} - |A|^2 \, \mathcal{C}(\mathbf{q}_{2}^{i} ) \mathbf{q}_1^{r} &=& 0 \nonumber \\
- \omega \mathcal{M} \mathbf{q}_1^{i}  + \mathcal{L}\left[\mathbf{q}_0 - |A|^2 \, \mathbf{q}_{2}^{r} \right]  \mathbf{q}_1^{r} - |A|^2 \, \mathcal{C}(\mathbf{q}_{2}^{i} ) \mathbf{q}_1^{i} &=& 0 \nonumber 
\end{eqnarray}
governing the real-valued variables $\mathbf{q}_{1}^{r}$ and $\mathbf{q}_{1}^{i}$. By gathering them into the (real-valued) extended vector $\mathbf{Q}_{1}=(\mathbf{q}_{1}^{r},\mathbf{q}_{1}^{i})^{T}$ and multiplying the second equation by $(-1)$, the above system is rewritten  
\begin{eqnarray}\label{eq:extendedfirstharmonic}
\left( \omega \mathcal{M}_{e} + \mathcal{L}_{e}(\mathbf{q}_{0},|A|^2 \mathbf{q}_{2}) \right)  \mathbf{Q}_{1} = \mathbf{0} 
\end{eqnarray}
where $\mathcal{M}_{e}= \begin{bmatrix}
   \mathcal{M} & 0 \\[2mm]
0 & \mathcal{M} \end{bmatrix}$ and $\mathcal{L}_{e}$ is the \textit{extended mean-flow} operator defined as 
\begin{eqnarray}
\mathcal{L}_{e}(\mathbf{q}_{0}, |A|^2 \mathbf{q}_{2}) = \begin{bmatrix}
-|A|^2 \, \mathcal{C}(\mathbf{q}_2^{i}) & \mathcal{L}\left[\mathbf{q}_0+|A|^2 \mathbf{q}_2^{r} \right]      \\[2mm]
-\mathcal{L}\left[\mathbf{q}_0-|A|^2 \mathbf{q}_2^{r}\right] & |A|^2 \, \mathcal{C}(\mathbf{q}_2^{i})  
 \end{bmatrix}
 \nonumber
\end{eqnarray}
that depends not only on the mean-flow $\mathbf{q}_0$ but also on the second harmonic $|A|^2 \mathbf{q}_2$.  Assuming they are both known (computed using \ref{eq:F-def} as explained before), the following generalized eigenvalue problem can be solved        
\begin{equation}\label{eq:extended-eigvalue}
\left[ \left( \omega_{e} + \rm{i} \sigma_{e} \right) \mathcal{M}_{e}  + \mathcal{L}_{e} (\mathbf{q}_0,|A|^2 \mathbf{q}_2)\right] \mathbf{\hat{Q}}_{e} = \mathbf{0}
\end{equation}
where $\mathbf{\hat{Q}}_{e}=(\mathbf{\hat{q}}_{e}^{1},\mathbf{\hat{q}}_{e}^{2})^{T}$ is a complex-valued eigenvector associated to a complex eigenvalue $\omega_{e} + \rm{i} \sigma_{e} $.  In the following, they will be called the extended eigenmode and the extended eigenvalue, respectively. Note that the real part of the extended eigenvalue corresponds to the physical frequency while the imaginary part  corresponds to the growth rate. Comparing \ref{eq:extended-eigvalue} with \ref{eq:extendedfirstharmonic}, one expects the existence of an extended eigenvalue, whose imaginary part is equal to zero ($\sigma_{e}=0$) while its real part is equal to the fundamental frequency ($\omega_{e}=\omega$), and the real corresponding eigenvector approximate the first harmonic ($\mathbf{\hat{Q}}_{e} = \mathbf{Q}_{1}$). This property of limit cycles is called the Extended Real Zero Imaginary Frequency (ERZIF) and can be sum up as
\begin{equation}
\sigma_{e} = 0 \;\;,\;\; \omega_{e} =\omega \;\;,\;\; \mathbf{\hat{q}}_{e}^{1} = \mathbf{q}_{1}^{r} \;\;,\;\; \mathbf{\hat{q}}_{e}^{2} = \mathbf{q}_{1}^{i}  
\end{equation}
Before applying this extended mean flow analysis to the circular cylinder flow configuration, it is worthwhile to brieflty mention one property of the extended eigenvalue problem \ref{eq:extended-eigvalue}. It can be shown that if $\omega_e+\rm{i} \sigma_{e}$ is an eigenvalue  associated to the eigenvector $\mathbf{\hat{Q}}_{e}$, then $\omega_e-\rm{i} \sigma_e$ is also an eigenvalue associated to the same eigenvector $\mathbf{\hat{Q}}_{e}$. The extended eigenvalue spectrum is therefore symmetric with respect to the  axis $\sigma_{e}=0$.
\section{Results}\label{sec:results}
\begin{figure}
\centering
\includegraphics[width=8cm]{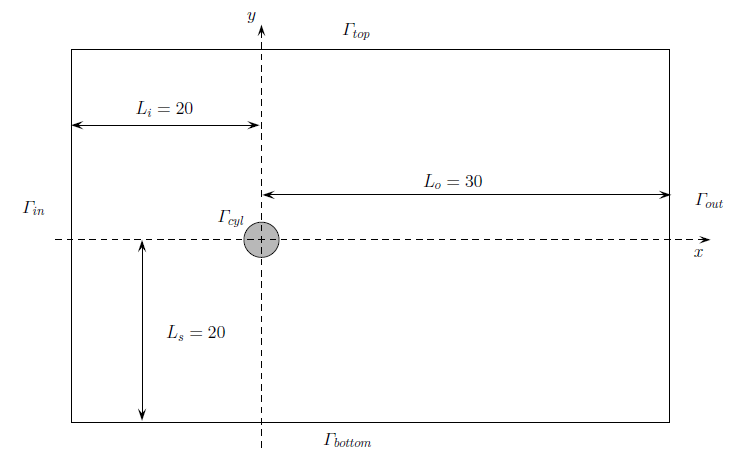} \\
\caption{Sketch of the computational domain $\Omega$ employed for the numerical simulation of the flow past a circular cylinder. All lengths are made non-dimensional using the cylinder diameter $D$.}
 \label{fig:cyl-sketch}
\end{figure}

\begin{figure}
\centering
\begin{tabular}{l}
(a) \\
\includegraphics[width=7cm]{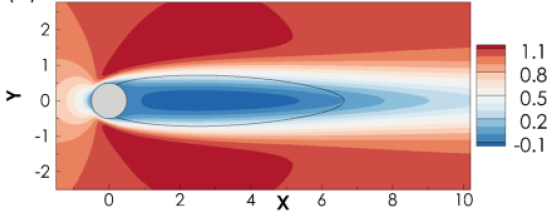} \\
(b)\\
\includegraphics[width=7cm]{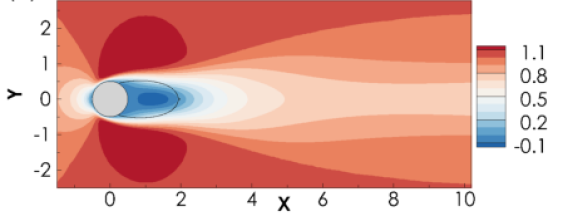} \\
(c)\\
\includegraphics[width=7cm]{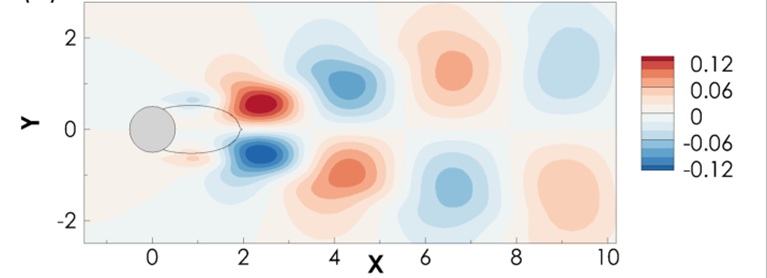} \\
(d)\\
\includegraphics[width=7cm]{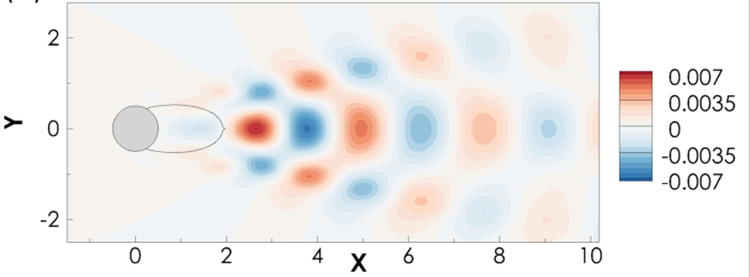} 
\end{tabular}
\caption{Streamwise velocity of (a) the base flow $\mathbf{q}_{b}$,  (b) the mean flow $\mathbf{q}_{0}$, (c) the real part of the first harmonic $\mathbf{q}_{1}^{r}$ and (d) the real part of the second harmonic $\mathbf{q}_{2}^{r}$ for the circular cylinder configuration at $\textit{Re}=100$.}
\label{fig:cyl-basemeanfirstsecond}
\end{figure}

\begin{figure}
\centering
\begin{tabular}{l}
(a) \\
\includegraphics[width=7cm]{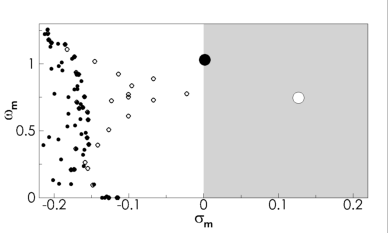} \\
(b) \\
 \includegraphics[width=7cm]{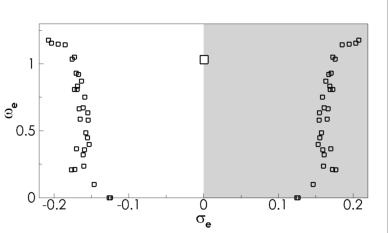} \\
(c) \\
\includegraphics[width=7cm]{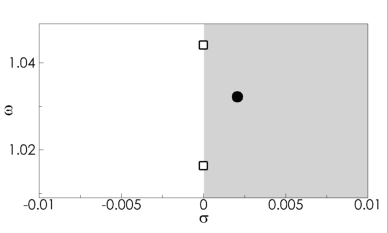} 
\end{tabular}
\caption{Eigenvalue spectra of (a) the base-flow $\mathcal{L}(\mathbf{q}_b)$ and mean-flow $\mathcal{L}(\mathbf{q}_{0})$ operators and (b) the extended mean flow operator $\mathcal{L}_{e}(\mathbf{q}_0,|A|^2 \mathbf{q}_2)$. (c)  Close-up view.  Base-flow and mean-flow eigenvalues are depicted with open and filled circles, respectively, while extended mean-flow eigenvalues are depicted with white squares.  $\textit{Re}=100$ }
\label{fig:cyl-eigs}
\end{figure}

The two-dimensional cylinder flow configuration is shown in \ref{fig:cyl-sketch} which indicates the lengths of the computational domain $\Omega$ in the streamwise $x$ and cross-stream $y$ directions. The nonlinear unsteady equations, nonlinear steady equations and eigenvalue problems introduced in the previous section are all solved using a finite-element method for the spatial discretization, as in \cite{marquet-sipp-jacquin-2008}. To solve the nonlinear unsteady or steady equations, a unit streamwise velocity profile is imposed at the inlet boundary $\Gamma_{in}$ while  the \textit{natural} outflow boundary condition $-p + \textit{Re}^{-1}\partial_{x} u=0$, $\partial_{x} v=0$ is imposed at the outlet $\Gamma_{out}$, where $\mathbf{u}=(u,v)^T$ are the velocity components in the considered reference coordinate system. On the top 
and bottom boundaries, $\Gamma_{top}$ and $\Gamma_{bottom}$, symmetry conditions $\partial_y u=0$, $v=0$ are used. For solving the eigenvalue problems, these  boundary conditions are applied in homogeneous form. The mesh is made of $47 000$ triangles, with a refinement region close to the body wall and in the cylinder near-wake, resulting in a minimum mesh size of $\Delta x\approx\Delta y\approx 0.02$. Unsteady direct numerical simulations are performed using the non-dimensional time steps of $\Delta t = 0.02$ for $\textit{Re}=60$, $\Delta t = 0.015$ for $\textit{Re}=100$ and $\Delta t = 0.01$ for $\textit{Re}=150$. \\
\\
The two-dimensional base-flow computed for $\textit{Re}=100$ is depicted in \ref{fig:cyl-basemeanfirstsecond}(a) with the streamwise velocity. The stability analysis of this base flow is performed by computing the eigenvalues with largest growth rates in \ref{eq:F-eigbaseflow}.  They are depicted with open circles in \ref{fig:cyl-eigs}(a). The large open circle, that lies in the gray plane, highlights the position of the unstable eigenvalue ($\sigma_b > 0$). Initializing the unsteady nonlinear simulations with a superposition of the base flow and a small-amplitude of the corresponding eigenmode (not shown here), a linear growth followed by a nonlinear saturation of the fluctuation is observed. The flow then settles down to a fully developed vortex-shedding regime characterized by a period $T=6.01$ corresponding to the fundamental frequency $\omega=1.0446$. 
Once the limit cycle is established and the flow period determined, the solution is advanced in time over one period to compute the mean flow as well as the first and second harmonics using \ref{eq:F-def}. The mean-flow shown in \ref{fig:cyl-basemeanfirstsecond}(b) displays a much shorter recirculation region than the base-flow. 
The first-harmonic, displayed in \ref{fig:cyl-basemeanfirstsecond}(c) with its real part,  is normalized using \ref{eq:normalization} where the spatial domain $\Omega_{s}$ extends in the range $-15 \le x \le 15$ and $-15 \le y \le 15$. The amplitude of the first harmonic is $A=1.5472$ for that Reynolds number.  \ref{tab:cyl-dns} reports the amplitude and frequency of the limit cycles obtained for other values of the Reynolds number.
\begin{table}[hbt]
\centering
 \begin{tabular}{|rcc|}
\hline 
$\textit{Re}$  \vline  & $|A|$ & $\omega$  \\ 
 \hline
   $60$  \vline & $1.13205$  & $0.8631$ \\ 
   $\mathbf{100}$  \vline 	 & $\mathbf{1.5472}$ & $\mathbf{1.0446}$ \\
   $150$ \vline  & $1.7932$  & $1.1657$ \\
\hline  
 \end{tabular}
 \caption{Amplitude and frequency of the first harmonic extracted from the time-periodic flows  computed for three values of the Reynolds number using unsteady numerical simulations.}
 \label{tab:cyl-dns}
\end{table}
The first harmonic exhibits an oscillating pattern in the streamwise direction and its streamwise velocity breaks the symmetry of the mean flow with respect to the $x$-axis. The spatial pattern of the second harmonic, shown in \ref{fig:cyl-basemeanfirstsecond}(d), also oscillates  in the streamwise direction, with a wavelength approximatively equal to half of the first-harmonic wavelength. The symmetry of the streamwise velocity is now similar to the one for the mean flow. \\

\begin{figure}
\centering
\begin{tabular}{l}
(a) \\
\includegraphics[width=8cm]{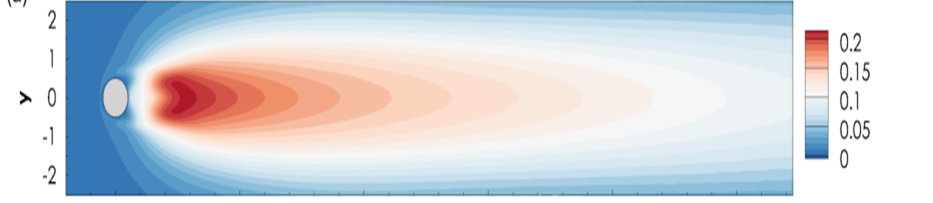} \\
(b) \\
\includegraphics[width=8cm]{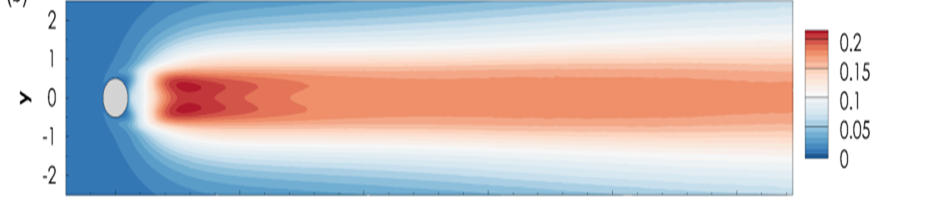} 
\end{tabular}
\caption{Spatial distribution of the velocity magnitude for (a) the first harmonic and  (b) the mean-flow eigenmode.}
\label{fig:cyl-meanflowmodes}
\end{figure}

The mean-flow eigenvalues, computed by solving \ref{eq:F-eigmeanflow} and first reported by Barkley \cite{barkley-2006} for that flow configuration, are shown in \ref{fig:cyl-eigs}(a) with black circles. The large black circle highlights the almost marginal eigenvalue. For $\textit{Re}=100$, its growth rate is $\sigma_m = 0.0020$ and its frequency $\omega_m=1.0322$. Values obtained for other values of the Reynolds number are reported in \ref{tab:cyl-eigs}. In all cases, the mean-flow frequency $\omega_{m}$ gives a much better approximation of the fundamental frequency $\omega$ than the base-flow frequency $\omega_b$. 
The spatial structures of the first harmonic and marginal mean-flow mode and first harmonic are shown in \ref{fig:cyl-meanflowmodes}(a) and (b), respectively. The magnitude of the velocity, equal to the square root of the mean kinetic energy, is displayed. In both cases, the largest amplitude is reached in the vicinity of the mean recirculation region. In the near-wake of the cylinder, the mean-flow eigenmode well approximates the first harmonic. In the far-wake of the cylinder, the amplitude of the mean-flow eigenmode does not decay, unlike what is observed for the first harmonic. We thus conclude that neglecting the quadratic interaction $(\mathcal{C}(\mathbf{q}_{2})\mathbf{q}_1^*)$  in the first-harmonic equation does not impact 
neither the frequency nor the near-wake of the fluctuation, but changes the spatial decay in the far-wake.\\
\\
The extended eigenvalue problem \ref{eq:extended-eigvalue} is solved using the mean-flow and the second-harmonic previously shown for $\textit{Re}=100$. We are now interested by determining real eigenvalues. Indeed, complex eigenvalues of the extended problem write $\omega_{e} + \rm{i} \sigma_{e}$, where their real part corresponds to a physical frequency and their imaginary part to a physical growth rate. The extended eigenvalue spectrum is depicted in \ref{fig:cyl-eigs}(b). First, we observe that the spectrum is symmetric with respect to the axis $\sigma_{e}=0$, meaning that when $\omega_{e} + \rm{i} \sigma_{e}$ is an eigenvalue, then $\omega_{e} - \rm{i} \sigma_{e}$ is also an eigenvalue. This is due to the structure of the extended operator, as noticed in the previous section. Secondly, the large white rectangle highlights the position of an eigenvalue close to the zero growth rate axis. The close-up view around that position displayed  in \ref{fig:cyl-eigs}(c) reveals that 
there are two eigenvalues lying exactly on the axis (zero growth rate) and having close but different frequencies. For comparison, the position of the marginally stable mean-flow eigenvalue is also reported in the figure with the black circle. 
\begin{table}
\centering
 \begin{tabular}{rcrcr}
& \multicolumn{2}{c}{Base flow} & \multicolumn{2}{c}{Mean flow} \\ 
\hline
$\textit{Re}$  \vline  & $\sigma_b$ & $\omega_b$  \vline & $\sigma_m$ & $\omega_m$  \\ 
\hline 
   $60$  \vline  & $0.0487$ & $0.7578$ 	 \vline        & $-0.0001$ & $0.8571$  \\ 
   $\mathbf{100}$ \vline  & $\mathbf{0.1257}$ & $\mathbf{0.7389}$   \vline        & $\mathbf{0.0020}$ & $\mathbf{1.0322}$   \\
   $150$ \vline  & $0.0503$ & $0.7390$   \vline       & $0.0006$ & $1.1526$   \\
\hline \\
 & \multicolumn{4}{c}{Extended Mean flow} \\ 
\hline 
$\textit{Re}$  \vline  & $\sigma_e$ & $\omega_e$ \vline & $\sigma_e$ & $\omega_e$    \\ 
   \hline
   $60$  \vline & $0$ & $0.8507$  \vline  & $0$ & $0.8631$ \\ 
   $\mathbf{100}$  \vline 	 &  $\mathbf{0}$ & $\mathbf{1.0164}$ \vline  & $\mathbf{0}$ & $\mathbf{1.0441}$ \\
   $150$ \vline  & $0$ & $1.1318$ \vline & $0$ & $1.1637$  \\
\hline
\end{tabular}
 \caption{(Top) Growth rate and frequency of the leading eigenvalues obtained for the base flow and mean flow analysis. (Bottom) Growth rate and frequency of the two  (real) eigenvalues obtained with the extended mean flow analysis for three values of $\textit{Re}$.}
 \label{tab:cyl-eigs}
\end{table}
The exact values obtained for the growth rate and frequency of these two eigenvalues are reported in \ref{tab:cyl-eigs} for the three values of Reynolds number investigated. For the growth rates $\sigma_{e}$, we systematically report the value $0$ since they are lower than $10^{-11}$ in all cases. For the frequency, it is interesting to note that one extended eigenvalue gives a very  good approximation ($\omega_{e} = 1.0441$, right column) of the limit-cycle frequency ($\omega=1.0446$ for $\textit{Re}=100$),  while the other one ($\omega_{e} = 1.0164$, left column) gives a poorer approximation compared to the mean flow frequency ($\omega_{m}=1.0322$). \\
 
\begin{figure}
\begin{tabular}{l}
(a) \\
\includegraphics[width=8cm]{./Figures/MagnitudeFirstHarmo.png} \\
(b) \\
\includegraphics[width=8cm]{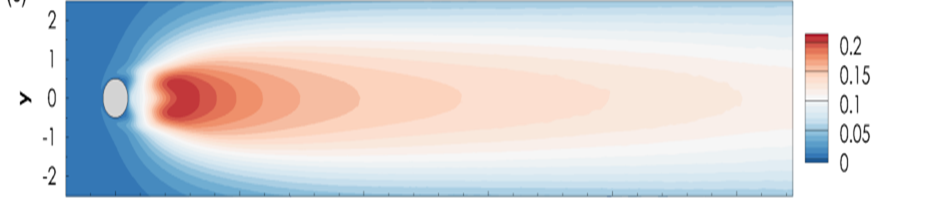} \\
(c) \\
\includegraphics[width=8cm]{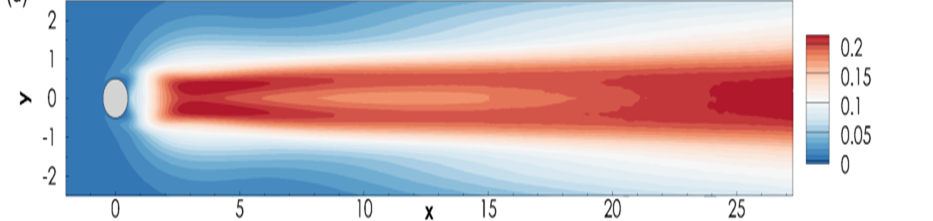} 
\end{tabular}
\caption{Extended mean-flow eigenmodes. Spatial distrution of the velocity magnitude for (a) the first harmonic 
and the eigenmodes oscillating at (b) $\omega_{e}=1.0441$ and (c)   $\omega_{e}=1.0164$ for $\textit{Re}=100$.}
\label{fig:cyl-extendedmeanflowmodes}
\end{figure}

Let us now examine the spatial structure of the extended eigenmodes associated to these two eigenvalues, and called 
the high-frequency and low-frequency extended modes. They are respectively displayed in \ref{fig:cyl-extendedmeanflowmodes}(b) and (c) by means of the magnitude of the velocity field. For comparison, the first harmonic is reproduced in \ref{fig:cyl-extendedmeanflowmodes}(a). A substantial difference is observed in the shape of the two modes. Consistently with the above results, the high-frequency mode (\ref{fig:cyl-extendedmeanflowmodes}-b) is found to better approximate the shape of the first harmonic (\ref{fig:cyl-extendedmeanflowmodes}-a). In particular, the spatial decay of the velocity in the far-wake of the cylinder is well reproduced by the high-frequency mode, unlike the low-frequency mode (\ref{fig:cyl-extendedmeanflowmodes}-c) which even exhibits a spatial growth of the velocity in the far-wake.  The same tendency has been observed for all values of the Reynolds number investigated.  To select between the two extended modes without knowing the structure of the first harmonic, \ref{eq:F-eq:2} governing the second-harmonic can be solved by replacing (in the right-hand side term) the first-harmonic with the spatial structure given by each extended mode. Results are not shown here but indicate that, consistently with the above results, the solution obtained with the high-frequency extended mode is close to the second-harmonic, unlike the one obtained with the low-frequency extended mode. \\

\section{Conclusions}
The extended mean-flow analysis has been introduced in this paper to analyze limit cycles. It is based on an eigenvalue analysis of an extended operator, that appears naturally when reformulating the equation of the first-harmonic of the time-periodic flow, and that depends not only on the mean-flow but also on the second-harmonic. Their effects are 
thus both taken into account, unlike the mean-flow analysis. This analysis has been  applied to the circular cylinder flow configuration and revealed the existence of two eigenmodes whose growth rate is exactly equal to zero. One of this eigenmode oscillates at a frequency almost equal to the frequency of the time-periodic flow, and its spatial structure exhibits a decay of energy observed in the first-harmonic, but not in the eigenmode obtained with the mean-flow analysis. 

\section*{Acknowledgment}
This project has received funding from the European Research Council (ERC) under the European Union Horizon H2020 research and innovation programm (grant agreement number 638307). 

\bibliography{references} 
\bibliographystyle{ieeetr}
\end{document}